\documentclass[twocolumn,amsmath,amssymb,aps,prstab,floatfix]{revtex4-1}
\usepackage[latin9]{inputenc}
\usepackage{color}
\usepackage{amstext}
\usepackage{amssymb}
\usepackage{textcomp}
\usepackage{graphicx}
\usepackage{soul}
\usepackage{epstopdf}
\usepackage{wasysym}
\usepackage[version=3]{mhchem}
\newcommand*{\rom}[1]{\expandafter\@slowromancap\romannumeral #1@}
\usepackage{pgfplots}

\usepackage{moreverb}
\makeatletter
\@ifundefined{textcolor}{}
{%
	\definecolor{BLACK}{gray}{0}
	\definecolor{WHITE}{gray}{1}
	\definecolor{RED}{rgb}{1,0,0}
	\definecolor{GREEN}{rgb}{0,1,0}
	\definecolor{BLUE}{rgb}{0,0,1}
	\definecolor{CYAN}{cmyk}{1,0,0,0}
	\definecolor{MAGENTA}{cmyk}{0,1,0,0}
	\definecolor{YELLOW}{cmyk}{0,0,1,0}
}

\makeatother

\begin{document}
	
	\title{Manipulation of Viscous Fingering in a Radially-Tapered Cell Geometry}
	
	\author{Gr\'{e}goire Bongrand}
	
	\author{Peichun Amy Tsai$^\ast$}

	\affiliation{Department of Mechanical Engineering, University of Alberta,\\ Edmonton, Alberta, Canada T6G 2G8}

	\begin{abstract}
	When a more mobile fluid displaces another immiscible one in a porous medium, viscous fingering propagates with a partial sweep, which hinders oil recovery and soil remedy. 
	We experimentally investigate the feasibility of tuning such fingering propagation in a non-uniform narrow passage with a radial injection, which is widely used in various applications.
	We show that a radially converging cell can suppress the common viscous fingering observed in a uniform passage, and a full sweep of the displaced fluid is then achieved. The injection flow rate $Q$ can be further exploited to manipulate the viscous fingering instability. For a fixed gap gradient $\alpha$, our experimental results show a full sweep at a small $Q$ but partial displacement with fingering at a sufficient $Q$. Finally, by varying $\alpha$, we identify and characterize the variation of the critical threshold between stable and unstable displacements. Our experimental results reveal good agreement with theoretical predictions by a linear stability analysis.

	\end{abstract}
	
	\maketitle

The process of fluid-fluid displacement in a porous medium is omnipresent in nature and occurs in numerous applications~\cite{Hill1952}, for example, printing~\cite{pearson1960, pitts1961, taylor1963,braun1984, ruschak1985, rabaud1990}, groundwater hydrology \cite{carman1956,marle1981}, contamination propagation in soils \cite{sahimi2011}, \ce{CO2} geological sequestration \cite{Huppert2014, Cinar2007} and enhanced oil recovery (EOR) technologies~\cite{Buckley1942, lake1989, Blunt1993}. Manifested in finger-shaped propagation, an interfacial instability emerges whenever a fluid of high mobility, $M=k/\mu$, characterized by the ratio of permeability $k$ to fluid viscosity $\mu$, pushes an immiscible one of low mobility. Because of its importance for an abundance of technologies, fingering instability of viscous fluids has been extensively studied. In particular, the classic Saffman-Taylor instability \cite{saffman1958} for both rectangular \cite{Chuoke1959,Wooding1976,Bensimon1986,Lenormand1988} and radial displacements \cite{Paterson1981,Chen1987, Homsy1987} in a Hele--Shaw cell, comprised of two parallel plates, separated by a small and uniform gap $h_0$ with $k=h_0^2/12$, is a  
convenient paradigm of a porous media flow with a homogeneous permeability~\cite{Homsy1987}.

\begin{figure}[b]
	\includegraphics[width=0.4\textwidth,keepaspectratio]{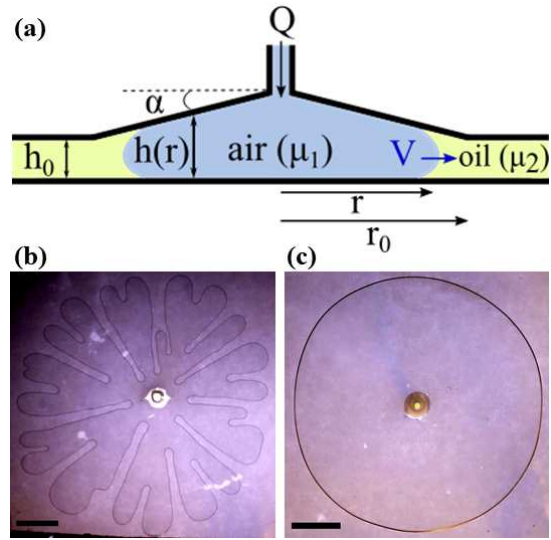}
	\protect\protect\caption{ (a) Schematic diagram of the side view of the experimental setup of immiscible fluid-fluid displacement in a radially converging passage, with a viscosity ratio $\lambda = \mu_2/\mu_1 = 8.8\times10^3$. (b) Snapshot of a classical viscous-fingering pattern obtained when air pushes oil in a flat radial Hele-Shaw cell with $h_0=1.2$ mm and $Q=40$ mL/min. (c) In contrast, snapshot of a stable interface with a complete sweep of oil by air in a radially tapered cell with $\alpha= -6.67 \times 10^{-2}$, $h_0=150$ $\mu m$, and $Q=40$ mL/min. For the experiments in (b) and (c), $h_0$ are chosen so that both configurations have equal fluid volumes. The scale bars in (b) and (c) are 2 cm.}
	\label{fig1} 
\end{figure}

\begin{figure*}[t]
\includegraphics[width=0.9\textwidth,keepaspectratio]{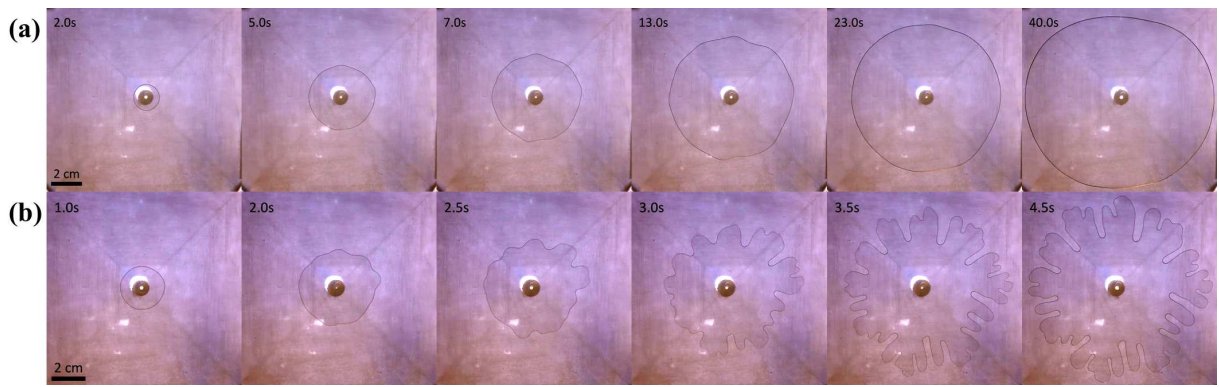}
\caption{(Supplemental Material) The dependence of sweep efficiency on flow rate $Q$: snapshots of two representative experiments \cite{SuppMat} of air displacing oil for the same geometrical configuration with the gap gradient $\alpha= -4.75 \times 10^{-2}$ and $h_0=500$ $\mu m$, but different flow rates: (a) stable displacement at $Q=40$ mL/min, whereas (b) viscous fingering at $Q=110$ mL/min.}
\label{fig2} 
\end{figure*}

Viscous fingering instabilities can be beneficial or detrimental depending upon the application. However, the control of viscous fingering has been a great challenge and hence investigated to a less extent since the mobility or viscosity contrast is often predetermined by the fluids chosen for the applications. Recently, such control has been achieved by controlling the injection rate of the displacing fluid~\cite{maxworthy2002,Li2009,Dias2012}, or using an upper elastic membrane forming a Hele-Shaw cell~\cite{Pihler-Puzovic2012,Pihler-Puzovic2013,Lister2013,Al-Housseiny2013a}. In addition,  a study has experimentally demonstrated the feasibility of suppressing fingering via a capillary effect using a rectangular Hele-Shaw cell with a converging gap~\cite{Al-Housseiny2012}, which has attracted renewed interest in the topic~\cite{Dias2013, Hazel2013, Setu2013, Kim2017, Diaz2017, Anjos2017,jackson2017}. Subsequently, several potential strategies for controlling viscous fingering instability have been explored, for instance, via wettability control of the fluids~\cite{Trojer2015, Zhao2016}, by experimentally lifting a plate with a time-dependent strategy at fixed flow rates \cite{Zheng2015} and numerically exploiting a gravitational (Rayleigh-Taylor) instability~\cite{Grenfell2017}. Nonetheless, systematic and thorough experimental investigation has yet to be carried out on the effect of depth gradients in a radial injection. In this Rapid Communications, we experimentally investigate viscous fingering problem in radial tapered Hele-Shaw cells [see Fig.~\ref{fig1}] and examine the impacts of depth gradients, radial propagation, and flow rates.

In our experiments, a variant of a Hele-Shaw cell with a converging gap is used to control the fingering instability with a radial injection [shown in Fig.~\ref{fig1}(a)]. The bottom plate is made of plexiglass and strengthened by another, thicker plate to avoid any bending.  The upper plate is tapered over a radius $r_0=7$cm, with a negative gap gradient, $\alpha = dh(r)/dr$, i.e., the ratio of the height to length of the tapered area. We control the height of the outer flat edge $h_0$, using translation stages with an accuracy of 10 $\mu$m.  The gap thickness inside the cell evolves linearly along the radius $h(r)=h_0 + \alpha(r-r_0)$, i.e., $\alpha = \frac{h_0-h(0)}{r_0} < 0$. The defending fluid is heavy mineral oil (viscosity $\mu_2$ =158 cP, Fisher Scientific), which initially is injected into and fully saturates the cell. We then inject air (viscosity $\mu_1 = 1.8 \times 10^{-5}$ Pa.s = $0.018$ cP) at a constant flow rate $Q$, ranging from 10 mL/min to 300 mL/min.

The observations are captured with a camera (Canon EOS 70D) at 30 fps (frames per second). 
We use ImageJ and Matlab to analyze the images and track the evolution of the interface. The local velocity, $V$, is calculated by tracking the interface position over a short period of time. We characterize the importance of the viscous forces relative to capillary forces using the capillary number, $Ca=12\mu_2V/\gamma$, where the surface tension of the oil, $\gamma=30$ mN/m, was measured using a tensiometer. The fluid combination in our experiments has a viscosity ratio  of $\lambda = \mu_2/\mu_1 = 8.8 \times 10^3$.

We first performed control experiments in flat and tapered cells, set with respective gap thicknesses so that both geometries have equivalent fluid volumes. As in the case of the classical Saffman-Taylor instability, when we conducted experiments with unfavorable viscosity-ratio displacement ($\mu_1 < \mu_2$) in a uniform cell, we observed unstable interfacial propagation with fingering, as shown in Fig.~\ref{fig1}(b). However, remarkably, when we carry out a similar experiment with $\mu_1 < \mu_2$ in a converging cell, the interface can be stabilized, as illustrated in Fig.~\ref{fig1}(c).

Through our experiments, we aim to understand the variation of the onset of instabilities for various flow configurations with different $\alpha$. By systematically varying the flow rate, $Q$ (for a fixed $\alpha$ and $h_0$), we observed a stable interface at low $Q$ throughout the experiment [see Fig.~\ref{fig2}(a)]. However, above a certain flow rate, the interface becomes wavy, and the instability grows as air displaces the oil, as shown in Fig.~\ref{fig2}(b). Physically, in the case of $M > 1$, the viscous pressure gradient gained is $\Delta P_\nu \sim \frac{\mu}{k}V$ and further destabilizes the interface as fluid travels radially outwards. For the radial injection, the interfacial velocity can significantly change due to mass conservation via a radially increasing cross-sectional area and a decreasing gap thickness, and hence alters $\Delta P_\nu$. On the other hand, the converging gap introduces a varying capillary pressure, $\Delta P_\gamma$, which increases and plays a crucial role in stabilizing the interface when fluids travel in a passage of decreasing depth.

\begin{figure}[t!]
	\includegraphics[width=0.45\textwidth,keepaspectratio]{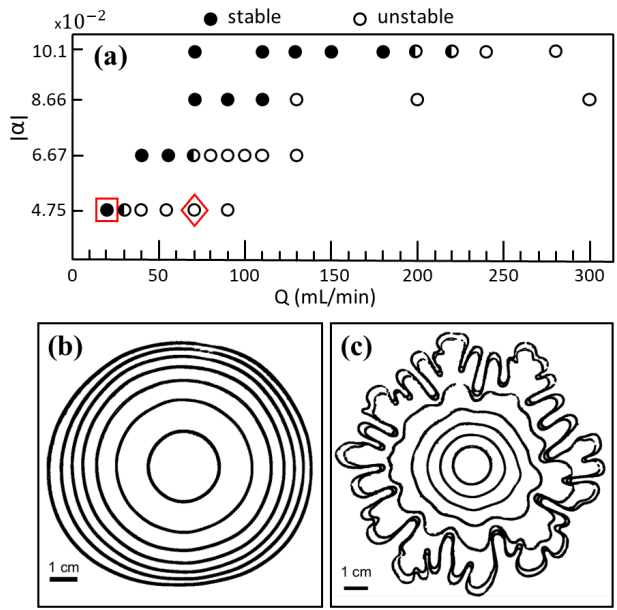}
	\protect\protect\caption{(a) Stability diagram of stable {\it vs.} unstable propagation front by varying the flow rate $Q$, for different gap gradients $\alpha$ (while $h_0=250$ $\mu m$). The general trend shows that for each $\alpha$, stable and complete sweep occurs at a relatively small $Q$ (denoted by $\bullet$), whereas unstable fingering propagation emerges at large $Q$ ($\circ$). 
For a radially-tapered cell of $\alpha= -4.75 \times 10^{-2}$, $h_0=250~\mu$m, time-evolution of top-view, stable interfaces with $Q=20$ mL/min in (b) (corresponding to \textcolor{red}{$\Box$} in (a)),  while unstable interfaces in (c) for $Q=70$ mL/min  (\textcolor{red}{$\Diamond$} in (a)). The time steps are $\Delta t=9$ s and $1$ s between each contours for (b) and (c), respectively.}
	\label{fig3} 
\end{figure}

By varying $Q$ gradually for different $\alpha$, we characterize the critical threshold of flow rate, $Q_c$, between stable [Fig.~\ref{fig3}(b)] and unstable [Fig.~\ref{fig3}(c)] displacements. Figure~\ref{fig3}(a) illustrates the phase diagram of stable {\it vs.} unstable interfacial propagation for different $\alpha$ and $Q$. This stability diagram reveals a general trend of increasing $Q_c$ with an increasing depth gradient, $|\alpha|$. 

\vspace{+0.1in}

Based on linear stability analysis, a theoretical model  to characterize the interface behaviour has been carried out 
for a rectangular \cite{Al-Housseiny2012} or radial fluid cell~\cite{Al-Housseiny2013}. The stability of a radially tapered viscous fingering interface is expressed in terms of its growth rate, $\sigma$~\cite{Al-Housseiny2013}: 
\begin{equation} 
\frac{r\sigma}{V}=-(1+\frac{\alpha r}{h}) + (1+\frac{2\alpha+ (h/r)^2}{Ca}) N - \frac{(h/r)^2}{Ca} N^3,\label{eqn1}
\end{equation}
where $N$ is the number of fingers, and $r$ is the radial position of the interface. $\sigma < 0$ characterizes a stable interface, while the critical transition occurs when $\sigma = 0$. We extract the values of $r$, $N$ and $V$ from the experiments showing the transitional behaviour from a stable to an unstable interface to analyze the growth rates. Under a constant flow rate, $Q$, as the fluid interface advances radially, $r=r(t)$, the interfacial speed $V$ changes due to mass conservation, i.e., $V=V(r(t))$. Consequently, the capillary number varies greatly with $r$.  For a gas displacing a wetting viscous liquid in a rectangular cell, the stability of the interface depends only on $Ca$ and $\alpha$, with a stable interface when $1+2\alpha/Ca<0$ \cite{Al-Housseiny2012}. On the other hand, the linear stability problem and the resulting growth rate, $\sigma$, for a radial variant Hele-Shaw cell are more complex than those in a rectangular configuration, due to the interplay between geometric and capillary parameters (e.g., $r$, $Ca$, and $N$). 

\begin{figure}[t!]
	\includegraphics[width=0.4\textwidth,keepaspectratio]{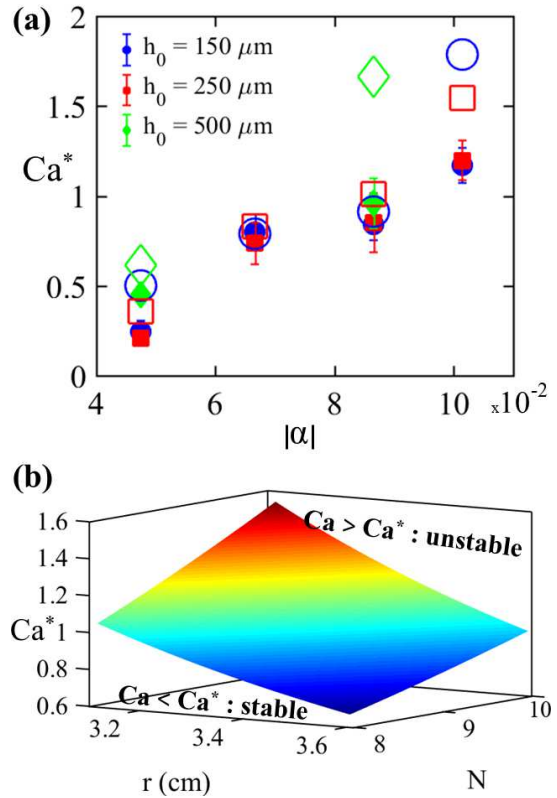}
	\protect\protect\caption{(a) Variation of the critical Capillary number $Ca^\ast$ separating stable {\it vs} unstable displacements for different depth gradients $\alpha$ and $h_0$. We compare the experimental values (\textcolor{blue}{$\CIRCLE$}, \textcolor{red}{$\blacksquare$}, \textcolor{green}{$\blacklozenge$}) to the theoretical $Ca^*_{th}$ (\textcolor{blue}{$\Circle$}, \textcolor{red}{$\square$}, \textcolor{green}{$\lozenge$}) derived from Eq.~(\ref{eqn2}) with $\alpha$, $r$ and $N$ from our experimental results and parameters. (b) Surface plot of theoretical $Ca^*_{th}$ greatly depends on $r$ and $N$ using Eq.~(\ref{eqn2}), for $\alpha= -8.66 \times 10^{-2}$ and $h_0=250$ $\mu m$, showing a stable displacement when $Ca < Ca^\ast$ whereas unstable one when $Ca > Ca^\ast$.}
	\label{fig4} 
\end{figure}

 To further compare our experimental results with the analytical prediction, we analyze the critical parameters at the transition when the wavy and fingering interface starts to set in.  The theoretical analysis of the growth rate $\sigma$ for our experimental conditions using Eq.~\ref{eqn1} is detailed in the Supplemental Material \cite{SuppMat}. Both results reveal a significant influence of $r$ and $N$ on the growth rate. We found that the growth rate strongly depends on $r$ for a certain critical fingering number, $N$, observed experimentally. As revealed in Fig.~S1,  a stable interface can become unstable within $1$ cm.  
Since multiple experimental parameters (e.g., $Q$, $\alpha$, $N$ and $r$) influence the growth rate of radial viscous fingering, we further analyze the critical Capillary number, $Ca^*$. 
Figure~\ref{fig4}(a) shows the comparison of experimentally observed $Ca^*$ with the theoretical prediction $Ca^*_{th}$ based on Eq.~\ref{eqn1} at the transition (i.e, by setting $\sigma=0$), where:
\begin{equation}
Ca^*_{th}=\frac{[2\alpha+(h/r)^2]N-(h/r)^2 N^3}{1+(\alpha r/h) -N}.\label{eqn2}
\end{equation}
The comparison, shown in Fig~\ref{fig4}(a), reveal consistent results between the experimental and theoretical critical capillary numbers. The experimental capillary number values, $Ca^*$, are determined via $Ca=12\mu_2V/\gamma$ for each $\alpha$ and $h_0$, where the local velocity, $V$, is measured with the videos showing a transitional displacement from a stable to an unstable interface. The theoretical predictions of $Ca^*_{th}$ are estimated using Eq.~\ref{eqn2}, with the inputs of $N$ and average $r$ analyzed from the same experimental videos.

Our experimental results show the small effect of $h_0$ on $Ca^\ast$ [\textcolor{blue}{$\CIRCLE$}, \textcolor{red}{$\blacksquare$}, \textcolor{green}{$\blacklozenge$} in Fig.~\ref{fig4}], while the theoretical critical $Ca^\ast_{th}$ from Eq.~\ref{eqn2} depends on $h(r) = h_0 + \alpha(r-r_0)$. Overall, a good agreement is found for smaller gap gradient. More importantly, the general trend of both theoretical $Ca^\ast_{th}$ and experimental $Ca^\ast$ increases with an increasing magnitude of the gap gradient. In other words, a larger interfacial velocity is required to trigger viscous fingering instability in a steeply convergent gap, wherecons large capillary pressure is present and acts to stabilize the interface.

For relatively larger $|\alpha|$, the deviation may be explained with the experimental difficulty of meeting the theoretical assumptions. The theoretical model Eq.~\ref{eqn1} \cite{Al-Housseiny2013} assumes a symmetric displacement, a constant static contact angle, and a small dimensionless depth variation, i.e., \(\lvert \lvert \frac{\alpha r}{N h} \rvert \rvert\)$<<1$. In contrast, our experiments have this dimensionless parameter ranging from $5 \times 10^{-2}$ to $10^{-1}$ and may have some surface roughness (due to the polish using fine sand papers) 
leading to inhomogeneous wetting and front propagation. These factors are likely the reasons for the deviation observed. In addition, the strong dependence of theoretical $Ca^*_{th}$ based on Eq.~\ref{eqn2} on $r$ and $N$ is shown in Fig.~\ref{fig4}(b). For instance, within a range of $0.5$ $cm$  in r and $N\pm1$, $Ca^*_{th}$ varies from 0.67 to 1.54 for the specific configuration presented in Fig.~\ref{fig4}(b). This drastic change within such a small radial position, $r$, and/or number of fingers, $N$, may explain the slight and moderate deviations observed for relatively moderate $|\alpha|$ in Fig.~\ref{fig4}(a).

In summary, we have experimentally demonstrated that the presence of a radial depth gradient (i.e., permeability variation), can alter significantly 
the viscous fingering instability and pattern. Using a converging passage, the classic viscous fingering commonly observed in a flat Hele-Shaw cell can be completely suppressed with a suitable flow rate. For each converging gradient, $\alpha$, we can tune the viscous-fingering instability from a stable to an unstable displacement by increasing the flow rate $Q$ injected. This critical flow rate is increased for a steeper gap gradient. We further compare our experimental results with a theoretical linear stability analysis, showing consistent dependence of the instability growth rate on radial location $r$ and $N$. From the experimental results with different $\alpha$, we further showed that the critical threshold $Ca^*$ increases with an increasing gap gradient $|\alpha|$, in good agreement with a recent theoretical prediction considering the effect of capillary pressure \cite{Al-Housseiny2013}. Our experimental results reveal for the first time the possibility of controlling interfacial instabilities with a radial injection in an inhomogeneous passage. The results of critical $Ca^\ast$ depending on $\alpha$, $r$, $h_0$ and $N$ are beneficial for the design and prediction of flow settings where the process of fluid-fluid displacement in a porous medium is crucial.
\\
\\{\bf Acknowledgments}~We gratefully thank G. M. Homsy for stimulating discussion.  P. A. T. acknowledges Natural Sciences and Engineering Research Council of Canada (NSERC) for the Discovery and Accelerator Grants as well as Canada Research Chair Program in Fluids and Interfaces.
\\
\\ $^\ast$ Email address of the corresponding author: 
\\P. A. Tsai (peichun.amy.tsai@ualberta.ca).

\bibliography{VF}
\end{document}